\documentclass[10pt,preprint,a4paper]{aastex}

\usepackage{amsmath}                
\usepackage{amsfonts}               
\usepackage{amssymb}                
\usepackage{epsfig}                 
\usepackage[left=2.1cm,top=2cm,right=1.8cm,nofoot]{geometry}

\newcommand{\yr}{{~\rm yr}}

\title{MERGER BY MIGRATION AT THE FINAL PHASE OF COMMON ENVELOPE EVOLUTION}

\author{Noam Soker\altaffilmark{1}}

\altaffiltext{1}{Department of Physics, Technion -- Israel Institute of
Technology, Haifa 32000 Israel; soker@physics.technion.ac.il}

\begin{document}

\begin{abstract}
I find the common envelope (CE) energy formalism, the CE $\alpha$-prescription, to be inadequate to predict the final
orbital separation of the CE evolution in massive envelopes.
I find that when the orbital separation decreases to $\sim 10$ times the final orbital separation predicted by
the CE $\alpha$-prescription,
the companion has not enough mass in its vicinity to carry away its angular momentum.
The core-secondary binary system must get rid of its angular momentum by interacting with mass further out.
The binary system interacts gravitationally with a rapidly-rotating flat envelope, in a
situation that resembles planet-migration in protoplanetary disks.
The envelope convection of the giant carries energy and angular momentum outward.
The basic assumption of the CE $\alpha$-prescription, that the binary system's gravitational
energy goes to unbind the envelope, breaks down.
Based on that, I claim that merger is a common outcome of the CE evolution of AGB and red super-giants stars
with an envelope to secondary mass ratio of $M_{\rm env}/M_2 \ga 5$.
I discuss some other puzzling observations that might be explained by the migration and merger processes.
\end{abstract}



\section{INTRODUCTION}
\label{sec:introduction}

The common envelope (CE) process is in the heart of the formation of many close binary systems
(e.g., \citealt{IbenLivio1993}; \citealt{TaamSandquist2000}; \citealt{Podsiadlowski2001}; \citealt{Webbink2008}; \citealt{TaamRicker2010}).
During the CE phase the orbital separation decreases due to gravitational drag and tidal interaction (e.g., \citealt{IbenLivio1993, RickerTaam2012}).
The transfer of orbital energy and angular momentum to the envelope, as well as other possible energy sources, lead
to the ejection of the envelope.
One of the major unsolved questions of the 36 years old (\citealt{Paczynski1976, vandenHeuvel1976}) CE process is the final orbital separation.

In the commonly used energy formalism of the CE the gravitational energy released by the spiraling-in binary system $E_G$, is
equated to the envelope binding energy (e.g., \citealt{Webbink1984, TaurisDewi2004}), $E_{\rm bind}$, with an efficiency $\alpha$: $\alpha E_G = E_{\rm bind}$.
This is termed the CE $\alpha$-prescription.
It is now a common practice to include the internal energy of the envelope in calculating the binding energy
(e.g. \citealt{Han1994, Maxted2002, Zorotovic2010, XuLi2010, Davis2011}), and some authors argue
for observational support for that (e.g., \citealt{Rebassa-Mansergas2012}).
\cite{IvanovaChaichenets2011} suggest that the enthalpy rather than the internal energy should be included in calculating the binding energy.

The basic assumption of the CE $\alpha$-prescription is that the binding energy is channelled to eject the envelope in a uniform manner
(for a thorough discussion of this prescription and other aspects of the CE evolution see \citealt{Ivanova2012}).
Namely, there is no separation between envelope parts. This assumption was put into question by \citet{KashiSoker2011} in cases where
the final CE phase is a rapid process.
While some studies do suggest rapid final evolution, (e.g., \citealt{RasioLivio1996}; \citealt{LivioSoker1988}),
on the order of days to few weeks, others argue for a final phase that lasts for months
(e.g., \citealt{SandquistTaam1998, DeMarco2003, DeMarco2009, Passy2011, RickerTaam2012}).
In any case, \citet{KashiSoker2011} studied the rapid deposition of energy in the very inner regions of a massive envelope of
an asymptotic giant branch (AGB) star, and found that a substantial fraction of the ejected mass does not reach the escape velocity.
\citet{KashiSoker2011} used a self-similar solution and followed the blast wave propagation from the center of the AGB outwards.
They showed that ˜$\sim 1-10$ per cent of the
ejected envelope remains bound to the remnant binary system.
They further argued that the bound gas falls-back and forms a circumbinary disc around the post-CE binary system.
The interaction of the circumbinary disc with the binary system will reduce the orbital separation
much more than expected in the CE $\alpha$-prescription.
The smaller orbital separation favors a merger at the end of the CE phase or a short time after, while the core is still hot.
A different  mechanism for merger at the termination of the CE phase was suggested by \cite{IvanovaChaichenets2011}.

A support to the separation of the envelope to ejected and bound segments comes from numerical simulations as well, e.g., \cite{Lombardi2006}.
\cite{SandquistTaam1998} showed that some of the mass of the envelope can remain bound to one or both of the interacting stars.
They simulated a $5~\rm{M_{\odot}}$ AGB interacting with a $0.6~\rm{M_{\odot}}$ companion, and found that the
companion unbinds $\sim 1.55~\rm{M_{\odot}}$ ($\simeq 23$ per cent) of the AGB envelope.
They also obtained a differentially rotating thick disk or torus at intermediate stages of the CE evolution.
\cite{Soker1992} (see also \citealt{Soker2004}) had analytically obtained a similar thick disk structure.
In these cases a rapid merging is expected as well.
\cite{Passy2011} present more extreme results in their simulations.
They find that when the envelope is lifted away from the binary $\gtrsim 80$ per cent of the envelope remains bound to the binary.
In their simulations rotation of the AGB was not included, so the material that remains bound is probably overestimated.
They conclude that in some cases parts of the AGB envelope remain bound or marginally bound to the remnant post-AGB core.
\cite{DeMarco2011} suggest that envelope material which is still bound to the binary system at the end of the CE
will fall back onto the system and will form a circumbinary disk.
They suggest that such a disk might have some dynamical effects on the binary period.

\cite{RickerTaam2012} performed 3D numerical simulation of a low mass binary composed of a $1.05 M_\odot$ red giant and a $0.6 M_\odot$ companion.
Although the AGB envelope mass is much lower than the masses considered here, some results are relevant.
\cite{RickerTaam2012} find that during the rapid inspiral phase (the plunge-in phase; see \citealt{Ivanova2012})
only a fraction of $\sim 25 \%$ of the energy released by the spiraling-in
process goes toward ejection of the envelope.
The mass loss rate at the end of their simulation is $\sim 2 M_\odot \yr^{-1}$.
As mentioned, \citet{Passy2011} find that most of the envelope stays bound at the end of their simulations
(on the comparison between the simulations of \citealt{Passy2011} and \citealt{RickerTaam2012} see \citealt{Ivanova2012}).
These results might imply that for a white dwarf (WD) companion of mass $\sim 0.6 M_\odot$ inside an AGB
star of mass $>4 M_\odot$, the envelope can stay bound for few years after the orbital separation has shrunk to several solar radii.
At the end of the simulation the envelope is concentrated around the equatorial plane of the binary system.
This result of a rapidly rotating disk-like envelope structure is one of the motivation for the present paper.

In the present paper I put into question some basic assumptions of the CE $\alpha$-prescription.
I do not dispute that the CE helps in ejecting the envelope.
However, most of the gravitational energy is released near the final expected separation.
There, I argue in section \ref{sec:viscous}, some assumptions
of the CE $\alpha$-prescription break down. Despite that, the CE $\alpha$-prescription might give a crude estimate of the final separation
in cases that don't end up in merger.
The final migration to merger is discussed in section \ref{sec:migra}, and a phenomenological toy model is proposed in section \ref{sec:scenario}.
Summary and implications for some puzzling observations are discussed in section \ref{sec:summary}.

\section{INTERACTION DEEP IN THE COMMON ENVELOPE}
\label{sec:viscous}
The density profile of AGB stellar envelopes can be approximated by $\rho(r) \propto r^{-\beta}$ where $\beta \simeq 2-2.4$.
For the purpose of this paper it is adequate to take for the envelope density
\begin{equation}
\rho(r)=\rho_0 \left(\frac{r}{r_0} \right)^{-2.2} =
\frac{0.8}{4 \pi} \frac{M_{\rm env}}{R_\ast^3} \left(\frac{r}{R_\ast} \right)^{-2.2},
\label{eq:rho1}
\end{equation}
where $\rho_0$ is the density at a reference radius $r_0$, $M_{\rm env}$ is the envelope mass, and $R_\ast$ is the stellar radius.
The envelope mass inside radius $r$ is given by
\begin{equation}
M_{\rm er}(r)= {M_{\rm env}} \left(\frac{r}{R_\ast} \right)^{0.8}=
0.33 \left(\frac{r}{10 R_\odot} \right)^{0.8}
\left(\frac{M_{\rm env}}{5 M_\odot} \right)
\left(\frac{R_\ast}{300 R_\odot} \right)^{-0.8} M_\odot.
\label{eq:menv1}
\end{equation}
The binding energy of the envelope is given by the expression
\begin{equation}
E_{\rm bind} = 0.5 \int_{r_m}^{R_\ast} \frac {G[M_{\rm core} + M_{\rm er}(r)]}{r} 4 \pi r^2 \rho(r) dr,
\label{eq:ebind1}
\end{equation}
where the factor 0.5 comes from the usage of the virial theorem and $M_{\rm core}$ is the core mass.
Because the integration diverges at low radii for the simple model used here,
the binding energy is calculated for the envelope residing above radius $r_m \sim 1 R_\odot$.
Performing the integration then gives the binding energy
\begin{equation}
\begin{split}
E_{\rm bind} (r_m) =
\frac{G M_{\rm env} M_{\rm core}}{R_\ast}
\frac{2}{3}
\left\{ 3 \left[ \left( \frac{R_\ast}{r_m} \right)^{0.2} -1 \right]
+
\frac{M_{\rm env}}{M_{\rm core}} \left[ 1 - \left( \frac{R_\ast}{r_m} \right)^{-0.6} \right] \right\}
\equiv \frac{G M_{\rm env} M_{\rm core}}{R_\ast} \Gamma,
\end{split}
\label{eq:ebind2}
\end{equation}
where the second equality defines $\Gamma$.
For a typical values of $M_{\rm env} \simeq   5 M_\odot$, $M_{\rm core} \simeq 0.7 M_\odot$, $R_\ast \simeq 300 R_\odot$ and $r_m \simeq 1 R_\odot$,
we find $\Gamma \simeq 9$.

According to the CE $\alpha$-prescription the final orbital separation would be determined by the equality
$0.5 \alpha G M_{\rm core} M_2/a_\alpha = E_{\rm bind}$, where $M_2$ is the secondary mass and $a_\alpha$ the final separation according
to the $\alpha$ prescription.
This gives
\begin{equation}
\frac{a_\alpha}{R_\ast}=0.01 \alpha
\frac{M_2}{0.2M_{\rm env}}
\left( \frac{\Gamma}{10} \right)^{-1}.
\label{eq:aalpha}
\end{equation}
Substituting $a_\alpha$ in equation (\ref{eq:menv1}) gives the envelope mass inner to that radius
\begin{equation}
\frac{M_{\rm er}(a_\alpha)}{M_2}= 0.13  \alpha^{0.8} 
\left( \frac{M_2}{0.2M_{\rm env}} \right)^{-0.2}
\left( \frac{\Gamma}{10} \right)^{-0.8}.
\label{eq:Malpha}
\end{equation}

For this scaling we find that $M_{\rm er}(2a_\alpha)=0.22M_2$ and $M_{\rm er}(3a_\alpha)=0.3M_2$,
and that $M_{\rm er}(r)=M_2$ at $r=13 a_\alpha$.
This result implies the following for CE inside evolved red giant stars (AGB, RGB, etc.).
As the orbital separation shrinks to $\sim 10 a_\alpha$, the companion has not enough mass in
its vicinity to carry away its angular momentum. Hence, the interaction with the gas in the vicinity of the companion
cannot cause much further spiraling-in, and no gravitational energy is liberated to unbind the envelope.
At this point the so called plunge-in phase, where the spiraling is dynamical and rapid  (see \citealt{Ivanova2012}), must end.
The core-secondary binary system must get rid of its angular momentum by interacting with mass residing at $r>a$,
where $a$ is the orbital separation at the given moment.
In other words, local interaction with the envelope gas at a typical distance of the Bondi-Hoyle-Lyttleton accretion radius
and tidal interaction with the mass residing at $r<a$, cannot cause further spiralling-in.
The binary system interacts gravitationally with a rapidly-rotating flat envelope \citep{RickerTaam2012}, where flat envelope refers to
a highly oblate envelope structure.
This situation becomes more like the planet-migration process in protoplanetary disks rather than the hydrodynamical
interaction of a secondary star in the classical prescription of a CE.
At this stage the flat envelope dissipates the energy deposited by the spiraling-in binary system, and creates heat that
can be transferred outward by convection.
The basic assumption of the CE $\alpha$-prescription, that the binary systems gravitational
energy goes to unbind the envelope, breaks down.

This discussion raises questions about the entire CE $\alpha$-prescription. Problems for the CE $\alpha$-prescription in explaining observations
were noted before, e.g., \cite{NelemansTout2005}.
\cite{NelemansTout2005} find the CE-$\gamma$ prescription to better fit observations.
In the CE$-\gamma$ prescription \citep{Nelemansetal2000} the angular momentum released by the spiraling-in binary
system is assumed to be carried away by mass loss.
The CE$-\gamma$ prescription also seems to break-down at $a \la 10 a_\alpha$, for the same reason the CE$-\alpha$ does:
there is not enough mass in the vicinity of the secondary and inward to transfer the angular momentum to.
In any case, the CE$-\gamma$ prescription has other severe problems \citep{Ivanova2012}.

\section{MIGRATION-INDUCED MERGER}
\label{sec:migra}

The conclusion from the above discussion is that when the orbital separation inside RGB and AGB stars decreases to a
radius where $M(a) \simeq M_2$,
which occurs at $a \gg a_\alpha$, the spiraling-in process substantially slows down.
Here as before, $a$ is the binary separation and $a_\alpha$ is the
final separation predicted by the CE $\alpha$-prescription.
Although numerical simulations do not always find rapid envelope rotation or a very flattened envelope (e.g., \citealt{Passy2011}),
I assume here that after
the slowing down of the spiraling-in process, such a state will be achieved.
The binary system transfers angular momentum to the rapidly-rotating circumbinary flat envelope, and consequently the
average radius of the envelope expands, and the binary separation decreases and the eccentricity increases
(e.g., \citealt{Artymowicz1991}).
I assume that the structure of the flattened envelope crudely resembles the structure of the disk studied in \cite{Artymowicz1991}.
In the calculations of \cite{Artymowicz1991} the disk extends from the nearest stable circumbinary orbit of $\sim 2.5 a$
up to $6a$.
I assume that the binary orbit is circular ($e_0=0$).
Scaling the results of \cite{Artymowicz1991} by values appropriate for the present study gives
the rate of change of the semi-major axis
\begin{equation}
\frac{\dot{a}}{a} \simeq -4.5 \times 10^{-5} \left(\frac{q_d}{0.1}\right) \Omega_b,   
\label{eq:adotoveraArty}
\end{equation}
where
\begin{equation}
\Omega_b=\sqrt{\frac{G(M_{\rm{core}}+M_2)}{a^3}} = 2.7
\left(\frac{a}{1 R_\odot} \right)^{-3/2}
\left(\frac{M_{\rm{core}}+M_2}{1.4 M_\odot} \right)^{1/2}  ~{\rm h}^{-1}
\label{eq:Omegab}
\end{equation}
is the binary orbital angular velocity, and
\begin{equation}
q_d=\frac{M_{\rm er}(6a)}{M_{\rm core}+M_2}. 
\label{eq:qd}
\end{equation}
Here $M_{\rm er}(6a)$ is the envelope mass inner to $r=6a$.
The spiraling-in timescale at $a=1 R_\odot$ with this scaling is
\begin{equation}
\tau_{\rm in} \equiv \frac{a}{\dot{a}} \simeq 1 \left(\frac{q_d}{0.1}\right)^{-1}
\left(\frac{a}{1 R_\odot} \right)^{3/2}
\left(\frac{M_{\rm{core}}+M_2}{1.4 M_\odot} \right)^{-1/2} \yr
\label{eq:timescale}
\end{equation}
The Keplerian (orbital) time decreases with decreasing separation, but then so does the mass in the flattened envelope.
The time scale of a year is long enough to allow the convective envelope to transport the heat outward.
Therefore, instead of expelling the entire envelope, the spiraling-in process increases the luminosity.
The wind mass loss rate might increase substantially, but a large portion of the envelope will stay bound in systems
where the initial envelope mass is $M_{\rm env} \ga {\rm several} \times M_2$.
The binary system will merge.

Equation (\ref{eq:adotoveraArty}), although crude for our case, none the less seems to be adequate for the present goals.
In a very recent paper \cite{ShiKrolik2012} performed 3D magnetohydrodynamic (MHD) simulations
of a circumbinary disk surrounding an equal mass binary system.
Similar to earlier studies, they find that strong torques by the binary clear a gap of radius $\sim 2 a$.
The binary system gains angular momentum by accretion from the disk, and loses it by the gravitational torque.
Over all the orbital separation decreases.
We note that the accretion rate onto the two stars in the present study is expected to be very low,
or practically zero, due to the high pressure around the core of the AGB star. To the contrary, mass is expected to flow outward.
 Although the torque found by \cite{ShiKrolik2012} is much larger than the one used in the $\alpha_d-$viscosity models
(not to be confused with the CE$-\alpha$), because of the opposite effect of accretion they find the spiraling-in
rate to be only a factor of $\sim 3$ higher than found here. The numerical coefficient in equation (\ref{eq:adotoveraArty})
according to \cite{ShiKrolik2012} is $-8 \times 10^{-5}$. The difference is of no significance for the goal
of the present paper.

For type II migration, where the planet is relatively massive and a gap is opened in the disk, the migration
rate is limited by viscous transport in the disk. The rate given by \cite{Alibert2005}, for example (their eq. 19),
gives a similar value to that in equation (\ref{eq:adotoveraArty}) here when the $\alpha_d-$viscosity coefficient is taken to be $\alpha_d \simeq 0.01-0.1$.
The same holds for the type II migration expression derived by \cite{Nelsonetal2000}, for example.

Over all, I conclude that the timescale for the final spiralling-in process for systems with
$M_{\rm env} \ga {\rm several} \times M_2$ is practically the
viscous time scale in the flattened envelope.
The major contribution to the viscosity is convection. This implies that on the same time scale that the binary system
spirals in, the convection transport the released gravitational energy out.
Instead of being used to expel the entire envelope, the released binary gravitational
energy is radiated away. The increase in luminosity will substantially expand the envelope and increase mass loss from the surface.
However, mass will stay in the vicinity of the binary system, and might lead to a merger (also \citealt{KashiSoker2011}).
A different process, based on the reexpansion of gas above the core and inner to the secondary orbit,  that might lead to merger was discussed
by \citet{Ivanova2011}. \citet{Ivanova2011}, as well as other papers dealing with post-CE merger, though, do use the energy formalism for the CE ejection
(the CE $\alpha$-prescription).
Here the fundamental assumptions of the CE $\alpha$-prescription are criticized.

\section{A PHENOMENOLOGICAL SCENARIO}
\label{sec:scenario}

 The termination of the CE phase with a core-secondary binary system interacting with material further out
is likely to be more violent than the migration of a planet (and more complicated).
Here I demonstrate a simple phenomenological approach as a demonstration of a plausible alternative to the CE $\alpha$-prescription
at the phase when the mass inward to the secondary is less than the mass of the secondary.
At this preliminary stage this should be considered more as a toy model rather than an established process.

I assume that bound mass is falling toward the binary system. No real disk is formed, but rather when the bound gas gets
to a distance of $\eta a$, where $ \eta \simeq 1$, from the binary system it acquires energy and escapes.
For simplicity consider the case where the core and secondary mass are about equal, $M_{\rm{core}} \simeq M_2$.
If the secondary mass is much below that of the core, the interaction will not be violent, and migration, more like planet
migration, takes place.
Consider that the core-secondary system interacts with bound gas of mass $M_b$, part of which fell to the center, at a typical distance $a$
from the center of mass. Each binary component is at a distance of $a/2$ from the center of mass.
As the core and secondary come closer to each other, so does the surrounding bound gas.
The binary system ejects the gas. Namely, the gas that is ejected carries a specific energy of
$d e_b \simeq G(M_{\rm{core}} + M_2)/(\eta a)$. The change in orbital separation is given by energy conservation
\begin{equation}
d \left(G \frac{M_{\rm{core}}M_2}{a} \right) \simeq \frac{ G({M_{\rm{core}}+M_2})} {\eta a} dM_b,
\label{eq:in1}
\end{equation}
where $M_b$ is the ejected mass from the binary vicinity (hence increases with time).
Integration of the left hand side from the orbital separation $a_b$ at the time the circumbinary interaction phase starts to a final orbital separation $a_f$,
and the right hand side from zero to $M_b$, yields, for $M_{\rm{core}} = M_2$,
\begin{equation}
a_f \simeq a_b e^{-4 M_b/\eta M_t}  =a_b (0.018)^{M_b/\eta M_t}
\label{eq:in2}
\end{equation}
where $M_t=M_{\rm{core}} + M_2$.

In this toy model the final orbital separation is very sensitive to the circumbinary bound mass present when the migration starts.
I emphasize again that this toy model applies only when the secondary mass is not much smaller than the primary mass.
For $a_b=10 R_\odot$ and $\eta=1$, for example, the final orbital separation in this toy model is $a_f=3.7 R_\odot$, $1.4R_\odot$, $0.5R_\odot$, and $0.2 R_\odot$,
for $M_b/M_t=0.25$, $0.5$, $0.75$, and $1$, respectively. The last two orbital separations might lead to a rapid merger.

\section{DISCUSSION AND SUMMARY}
\label{sec:summary}

In the previous sections I questioned the basic assumption of the common envelope (CE) energy formalism (the CE $\alpha$-prescription), and
found it to be inadequate to predict the final orbital separation of the CE evolution in massive envelopes.
The CE $\alpha$-prescription might help as long as the envelope mass inward to the location of the secondary is larger than the secondary mass.
This does not occur in the inner region when a massive envelope is involved, as evident from equation (\ref{eq:Malpha}).
Based on that, I claim that merger is a common outcome of the CE evolution of AGB and red super-giants stars
with an envelope to secondary mass ratio of $M_{\rm env}/M_2 \ga 5$. Many binary systems do survive of course.
Here $M_{\rm env}$ is the envelope mass at the beginning of the CE phase.

Mergers at the termination of the CE phase were discussed before (e.g., \citealt{IvanovaPod2003, Ivanova2011}), as well as the slowing down of the
spiraling-in process when the separation is small (see discussion in \citealt{IvanovaPod2003}). However, these authors and many others do
use the CE energy formalism.
In the energy formalism of the CE the binary gravitational energy that is released is equated to the envelope binding energy.
Most of the binary gravitational energy is released very deep in the envelope. It was found here that
where most of the gravitational energy is supposed to be released, there is not enough local envelope mass to absorb
the energy released by the binary system (for AGB stars with $M_{\rm env}/M_2 \ga 5$).
The energy must be absorbed by material residing further out in the flattened envelope, a process that proceeds on a viscous
time scale, as with migration of planets inside disks.
Viscosity is supplied by the convective envelope. The convective speed in AGB stars is close to the sound speed, and the same
convective motion
can transport the energy out. A large fraction of the released energy will go to increase the luminosity and inflate the envelope,
rather than to expel the inner parts of the envelope.

Merger at the termination of the CE phase might explain some puzzling observations. I list here several of these, each of which deserve much deeper study.
I do not list all possible outcomes of CE mergers, e.g., the formation of high-field Magnetic WDs \citep{Tout2008, Garcia2012} and NS-BH mergers
(e.g., \citealt{Dominik2012}), but rather limit myself to those scenarios that are theoretically less developed.

\emph{The Core-Degenerate (CD) scenario for Type Ia supernovae.}
\cite{SparksStecher1974} developed a merger scenario of a WD with an AGB core, in which the result is a type II SN.
In a later study \citet{LivioRiess2003} found that the merger of a WD with an AGB core can lead to a SN Ia that occurs
at the end of the CE phase or shortly after. Hydrogen lines will appear, and hence they attribute such a scenario to a rare type of SNe Ia.
\citet{KashiSoker2011} and \citet{IlkovSoker2012MNRAS} argue for a long time delay between the CE merger and the SN Ia explosion
due to a long spin-down process, and termed it the core-degenerate (CD) scenario for SN Ia.
They argue that this scenario might account for most SNe Ia.
An important conclusion of \citet{LivioRiess2003} and \citet{KashiSoker2011} is that for  merger to occur the AGB star should be massive.
As explained in \citet{Soker2011}, the CD scenario has some key differences from the double-degenerate scenario for SNe Ia, and hence should
be referred to as a separate scenario.

\emph{R Coronae Borealis stars with a massive expanding shell.}
The two leading models for the formation of R Coronae Borealis stars are a final-helium-shell flash of a post-AGB star and two WDs merger.
In a recent paper \cite{Clayton2011} find R Coronae Borealis to have signatures that are expected from both a WD merger and from a final-helium-shell flash
(see also \citealt{Lambert2011}).
The relatively high inferred mass of R Coronae Borealis and its high fluorine abundance support the merger model, while the massive, $\sim 2 M_\odot$, expanding shell
supports the final-helium-shell flash model.
I raise here the possibility that R Coronae Borealis and similar R CrB stars are formed from the merger of a WD with the core of an AGB star.

\emph{The bright edge of the planetary nebula luminosity function (PNLF).}
The bright edge of the PNLF in old and young stellar populations is the same (for a recent paper see \citealt{Ciardullo2012}).
Two scenarios were suggested to account for the puzzling finding that old stellar populations have as
bright planetary nebulae (PNs) as young stellar populations do.
One is that the bright part of the PNLF in old populations comes from blue-straggles \citep{Ciardullo2012}, namely, merger of main sequence stars.
The other one is that the brightest PNs in old populations are systems in transition from a symbiotic nebula phase to the PN phase
\citep{Soker2006}.
Here I raise the following scenario to account for bright PNs in old stellar population.
In a system of two low mass stars, the more massive lost its envelope to its companion while on the red giant branch (RGB). A helium WD of $\sim 0.2-0.3 M_\odot$ is formed.
Later, when the secondary becomes an AGB star, the WD remnant of the primary star merges with the AGB core before the hydrogen-rich envelope is lost.
The now more massive core can ionize the nebula to form a bright PN. This speculative scenario deserves more detailed study.

\emph{The formation of single sdB stars.}
Extreme horizontal branch star (EHB; also termed sdB or sdO stars) are horizontal branch stars that have lost most of their envelope at the end of the RGB phase.
Many sdB stars have a close stellar companion that went through CE phase. This explains the envelope ejection (e.g., \citealt{Han2003}).
However, many others might be formed through a CE phase with a substellar companion \citep{Soker1998}.
Although some substellar companions might survive the CE phase, there are recent indications that many of the substellar companions merge with the RGB core
\citep{Geier2011}.

\emph{Ultraluminous core collapse SNe.} There are hints that some ultraluminous core collapse SNe experience an extreme mass loss
episode $\sim 1-100 \yr$ before explosion (e.g., \citealt{Chomiuk2011}; see discussion in \citealt{Chevalier2012}).
I speculate that this coincidence can be caused by a companion that spirals-in inside the envelope and collides with the core.
The CE process causes the extreme mass loss rate, while the collision
with the core expedites the evolution toward explosion. The companion can be several solar masses main sequence star, or a neutron star.
{{{ A similar idea was put forward simultaneously by \cite{Chevalier2012}. }}}
This speculative scenario will be studied in a future paper.

\emph{Complicated planetary nebula structures.} The CE merger can lead to a PN with very complicated structures.
The PN NGC 5189, for example, has a very complicated structure \citep{Sabin2012},
and future studies should examine whether CE merger can indeed explain such a structure.

\emph{Missing long-orbital period post-CE binaries.}
The migration process at the termination of the CE phase reduces the orbital separation even if a merger does not occur.
This might explain the finding that all post CE binary systems in the homogeneous SDSS sample have orbital separation
well below expectation of the CE $\alpha$-prescription \citep{Rebassa-Mansergas2012}.

I thank Amit Kashi and an anonymous referee for helpful comments.
This research was supported by the Asher Fund for Space Research at the Technion and the Israel Science Foundation.

\end{document}